\documentclass[oldversion]{aa}     
\def \th {\thinspace}

\def \chisq {$\chi ^{2}$}
\def\approxgt{\mathrel{\hbox{\rlap{\lower.55ex \hbox {$\sim$}} \kern-.3em \raise.4ex \hbox{$>$}}}}
\def\lesssim{\mathrel{\hbox{\rlap{\lower.55ex \hbox {$\sim$}} \kern-.3em \raise.4ex \hbox{$<$}}}}
\def\approxlt{\mathrel{\hbox{\rlap{\lower.55ex \hbox {$\sim$}} \kern-.3em \raise.4ex \hbox{$<$}}}}
\def \degmark {^\circ}
\def \sun {\hbox {$\odot$}}
\usepackage{epsf}
\usepackage{graphicx}
\begin{document}

\title{A model for the Z-track phenomenon in GX\th 5-1 and observational evidence for
the physical origins of the kHz QPO}

\author{N. K. Jackson\inst{1}
\and M. J. Church\inst{1,2}
\and M. Ba\l uci\'nska-Church\inst{1,2}}
\institute{School of Physics and Astronomy, University of Birmingham,
           Birmingham, B15 2TT, UK\\
\and
          Astronomical Observatory, Jagiellonian University,
          ul. Orla 171, 30-244 Cracow, Poland.\\}
\offprints{njackson@star.sr.bham.ac.uk}

\date{Received 4 August 2008 ; Accepted 2 December 2008}
\titlerunning{The Z-track in GX\th 5-1}
\authorrunning{Jackson et al.}

\abstract{
We present the results of a combined investigation of the spectral and kHz QPO evolution around the Z-track 
in GX\th 5-1 based on high-quality {\it Rossi-XTE} data. In spectral analysis, we find that the Extended 
ADC emission model provides very good fits to all of the spectra, and the results point clearly to a model 
for the nature of the Z-track in this source, in agreement with previous results for the similar source 
GX\th 340+0. In this model, at the soft apex of the Z-track, the mass accretion rate $\dot M$ is at its minimum 
and the neutron star has its lowest temperature; but as the source moves along the normal branch, the 
luminosity of the Comptonized emission increases, indicating that $\dot M$ increases and the neutron star 
gets hotter. The measured flux $f$ of the neutron star emission increases by a factor of ten becoming 
super-Eddington, and we propose that this causes disruption of the inner disk and the formation of jets. 
In flaring, the luminosity of the dominant Comptonized emission from the accretion disk corona is constant, 
while the neutron star emission increases, and we propose for the first time that flaring consists of unstable 
nuclear burning on the neutron star, supported by the agreement between the measured mass accretion rate per 
unit area $\dot m$ at the onset of flaring and the theoretical critical value at which burning becomes unstable. 
There is a striking correlation between the frequencies of the kHz QPO and the ratio of the flux to the
Eddington value: $f/f_{\rm Edd}$, suggesting an explanation of the higher frequency QPO and of its variation 
along the Z-track. It is well known that a Keplerian orbit in the disk at this frequency corresponds to a 
position some distance from the neutron star; we propose that the oscillation always occurs at the inner disk 
edge, which moves radially outwards on the upper normal and horizontal branches as the measured increasing
radiation pressure increasingly disrupts the inner disk.
}
\maketitle

\section{Introduction}

The Z-track sources form the brightest group of low mass X-ray binaries (LMXB) containing a neutron 
star, with luminosities at or above the Eddington limit. They are characterised by having three 
distinct branches in a hardness-intensity diagram: the horizontal branch (HB), the normal branch (NB) 
and the flaring branch (FB) (Hasinger \& van der Klis 1989) showing that major physical changes take 
place within the sources but the nature of these has not been understood. It has been widely thought
that the physical and spectral changes are driven by change of a single parameter along the Z-track,
presumably the mass accretion rate (Priedhorsky et al. 1986), assumed to increase monotonically
in the direction HB - NB - FB. However, as discussed below, the evidence for this is rather limited.
Moreover the variation of the X-ray intensity does not obviously support this as the intensity 
does not increase monotonically in the direction HB - NB - FB, but decreases 
on the normal branch. Arguably the most important feature of the Z-track sources is the detection
of radio emission showing jets to be present but only in the upper normal and horizontal branches
(Penninx 1989). The presence of jets was dramatically demonstrated by extended radio observations
of the Z-track source \hbox{Sco\th X-1} (Fomalont et al. 2001) which revealed radio condensations moving
away from the source with velocity $v/c$ of 0.45. Thus the Z-track sources uniquely provide the
possibility of determining the physical conditions within the sources on the part of the
Z-track where radio is detected so telling us the conditions needed for the launching of jets.
Apart from this an understanding of the Z-track sources is essential to the basic understanding of 
LMXB in general. Extensive work has been carried out on the timing properties of Z-track sources (van 
der Klis et al. 1987; Hasinger \& van der Klis, 1989) and revealed the existence of quasi-periodic
oscillations (QPO) which change along the Z-track. However, analysis of the timing properties has not 
provided an explanation of the Z-track phenomenon. Spectral analysis is more likely to reveal the 
nature of the physical changes taking place as directly showing changes in the emission components 
during the spectral evolution along the Z-track, but spectral studies of the sources have been
hindered by lack of agreement over the emission model to be used.

\subsection{Spectral studies of the Z-track sources}

The nature of X-ray emission in LMXB has been controversial essentially because two radically
different types of model for the continuum are capable of describing the spectra in which a dominant 
Comptonized component and also a thermal component are clearly present. Analysis of the dipping class
of LMXB led to the ``Extended ADC'' model in which the dominant component is Comptonized emission from 
the accretion disk corona (ADC) plus thermal emission from the neutron star (Church \& 
Ba\l uci\'nska-Church 1995, 2004). Both spectral and timing analysis showed the Comptonizing ADC to be 
a region extending over a substantial fraction of the inner accretion disk (typically 15\%) and thin 
\hbox{($H/r$ $<<$ 1)} as discussed in more detail below. The other model is the Eastern model comprising 
multi-colour blackbody emission from the inner accretion disk plus Comptonized emission from a
central region close to the neutron star (Mitsuda et al. 1989). Previous spectral fitting of the 
Z-track sources has mostly been based on use of the Eastern model as summarized below. However, there is
now a body of evidence that the Comptonizing region is extended (Sect. 1.2) and is not a small central
region. Moreover, when the Eastern model was applied to the Z-track sources, it was difficult to
interpret the variation of spectral parameters and a physical explanation was often not given. 

An extensive investigation of spectral evolution along the Z-track in several sources using {\it Exosat}
data was carried out by Schulz et al. (1989) and by Schulz and Wijers (1993) using a model with neutron
star blackbody emission plus Comptonization. Hasinger et al. (1990) fitted Ginga data on Cyg X-2 with 
both Western and Eastern models. In the Western model the main change was an increase of blackbody temperature 
on climbing the normal branch. In the Eastern model, there was a systematic change in the Comptonization 
y-parameter on the normal branch. Because of the ability of both models to fit, it was not thought possible 
to distinguish between physical models. Consideration of multi-wavelength data and of the timing properties 
of the source led the authors to conclude that the mass accretion rate $\dot M$ increased monotonically in the 
direction HB - NB - FB although  X-ray intensity decreased on the NB. A possible explanation was that a 
thickening of the inner disk obscured the X-ray emission. The conclusion about $\dot M$ led Hasinger et al., 
following Hasinger (1988) and Lamb (1989), to propose a scenario in which $\dot M$ was small on the HB 
so that a thin disk existed together with a magnetosphere in which radio emission was generated. 
Increasing $\dot M$ resulted in the magnetosphere being enveloped by a vertically expanding disk at 
the hard apex, suppressing radio emission. At the soft apex $\dot M$ exceeds the Eddington value so that 
strong radiation pressure led to mass ejection above and below the disk.
Asai et al. (1994) fitted {\it Ginga} data on the NB and FB spectra of GX\th 5-1 with both Western and
Eastern models, but using the Eastern model in its original form consisting of disk blackbody
plus blackbody from the neutron star. This work concentrated on the emission feature found at 10 keV.

Spectral fitting of Cyg\th X-2 was carried out by Done et al. (2002) using the Eastern model assuming 
that the seed photons for Comptonization come from the neutron star or inner disk and so should be described 
by a blackbody spectrum. It was also assumed that the central Comptonizing cloud would illuminate the disk so
that reflection should be present (although in the Z-track sources the height of the inner disk is
much greater than that of the neutron star). The spectral fitting indicated an increasing flux of the 
disk blackbody in the lower normal and flaring branches where it was assumed that $\dot M$ increased; 
however, the disk temperature appeared constant so it was argued that the inner disk radius must increase. 
The Comptonized flux also remained constant when it might be expected to increase and it was not clear what 
caused the spectral parameters to vary in the way found. Agrawal \& Sreekumar (2003) used the Eastern model 
to fit {\it Rossi-XTE} spectra of GX\th 349+2 in the NB and FB using the form {\sc diskbb + comptt}, i.e. disk 
blackbody plus Comptonization with the seed photons having a Wien distribution. The inner disk temperature 
first increased on the NB then decreased in flaring; the inner disk radius first decreased then increased,
suggesting that the inner disk edge was moving to larger radial positions, and the authors proposed that 
there was thus an outflow of material . The electron temperature of the Comptonizing plasma first increased 
on the NB then decreased in flaring while the optical depth of the region first decreased then increased. 
It was suggested that the initial heating caused a density decrease and reduction in optical depth; on 
the FB the outflow proposed added density to the hot central corona causing the increase of optical depth 
on this branch.

Spectral fitting of broadband {\it BeppoSAX} data was carried out for GX\th 17+2, GX\th 349+2 and 
Cyg\th X-2 (di Salvo et al. 2000, 20001, 2002). 
In the case of Cyg\th X-2, the Eastern model was applied together with a power law to represent
a high energy tail, and it was argued that all three sources could be fitted by this model. In 
Cyg\th X-2 spectral evolution  consisted of an increase of disk blackbody temperature and decrease of blackbody 
radius moving from the HB to the NB, suggesting that the inner disk radius was shrinking although an 
explanation of this was not proposed. There were changes in temperature of the Comptonizing plasma 
and on the normal branch, the inferred radius of this region fell to 2 - 3 km, incompatible with the
neutron star being the source of seed photons unless there was a marked change in geometry 
from the normally assumed spherical Comptonizing cloud of the Eastern model.
D'Ai et al. (2007) investigated the Z-track in Sco\th X-1 using extensive numbers of observations with 
{\it RXTE}. The Eastern model in the form used by di Salvo et al. was used. The disk  blackbody 
temperature was highest on the flaring branch as expected on the standard assumption that $\dot M$ 
is largest here. The electron temperature of the Comptonizing plasma increased around the Z having
its highest values in flaring, while at the same time the optical depth decreased; however, the
reasons for these changes were not clear.

It is difficult to find consistency in the spectral fitting results and with the difficulty of 
interpreting the evolution of spectral parameters along the Z-track it is fair to say that there is 
no general concensus on the nature of the Z-track. Much effort was devoted to producing a theoretical 
model of of accretion in the inner disk in LMXB which proposed detailed physical changes taking place 
along the Z-track (Lamb 1989; Miller 1990;  Psaltis et al. 1995). This involves a magnetosphere at the 
inner accretion disk, and changes in the geometry and extent of the magnetosphere. The model assumes 
the main element of the Eastern model that is, a small central Comptonizing region. In the present work
we take the approach that the evidence favours an extended Comptonizing region and we test the
hypothesis that the Extended ADC model may provide a physically reasonable and consistent explanation
of the Z-track source GX\th 5-1.

\subsection{The extended ADC}

Our work on the dipping class of LMXB previously led to our proposal of the extended ADC model comprising 
blackbody emission from the neutron star plus Comptonized emission from an extended, thin ADC existing as 
a hot layer above the accretion disk. In the dipping sources with inclination angles between 65$\degmark$ 
and 85$\degmark$, reductions in X-ray intensity take place at the orbital period generally thought to involve
absorption in the bulge in the outer disk where the accretion flow impacts. Spectral evolution cannot be 
described in terms of absorption of a single emission component, but is more complex strongly constraining 
allowable models. It provides direct evidence for the extended nature of the Comptonized emission as this 
spectral component is removed only slowly during dipping showing the emitter to be extended. 
Secondly, the ADC size can be measured
directly by the technique of dip ingress timing since the ingress time depends on the size of the
major emission component: the Comptonized emission. This has shown that the radial extent of the ADC
is typically 50\th 000 km (Church \& Ba\l uci\'nska-Church 2004) also showing that the ADC, like the
disk is in general thin, with height $H$ at radial position $r$ having $H/r$ $<<$ 1. 
The extended ADC model was shown to give very good fits to the
complex spectral evolution in dipping in many observations of the dipping LMXB (Church et al. 1997,
1998a,b, 2005; Ba\l uci\'nska-Church et al. 1999, 2000; Smale et al. 2001; Barnard et al. 2001). In
addition, it was shown able to fit the spectra of all types of LMXB in a survey of LMXB made with
{\it ASCA} (Church \& Ba\l uci\'nska-Church 2001) and so was able to fit sources of all inclination angles.

The evidence favouring the extended ADC is thus strong and it is difficult to avoid the conclusion 
that the main assumption of the Eastern model is invalid.

The main difference between the spectral forms of the extended ADC model and the Eastern model
relates to the seed photons. In the Eastern model it is usually assumed that these originate on the neutron star
and can be described by a single temperature blackbody or by the Wien approximation. In the extended ADC model
the major source of seed photons must be the thermal emission of the disk below the ADC. Because of the 
temperature gradient $T(r)$ in the disk, the outer portions of the ADC will be above relatively cool parts 
of the disk providing a large population of soft seed photons ($kT$ $\sim$ 0.1 keV or less) and the seed 
photon spectrum is very different from a blackbody at $\sim$1 keV. Comptonization is well represented by a 
power law extending to energies lower than 0.1 keV, with a cut-off at high energies related to the maximum electron
energy (see Church \& Ba\l uci\'nska-Church 2004 for a full discussion). The use of a
simple blackbody for the neutron star emission in applying the extended ADC model reflects the general lack 
of evidence for modification of the blackbody in the atmosphere of the neutron star (Church et al. 2002). 
Taking into account many spectral studies of LMXB, including the spectra of X-ray bursts, the only evidence 
for modification is in a minority of bursts, with no evidence in non-burst emission. Church et al. showed that 
whether modification occurs depends critically on the electron density in the neutron star atmosphere and
that this is poorly known.

\subsection{A model for the Z-track}

We previously applied the extended ADC model to high quality {\it Rossi-X-ray Timing Explorer} data
from the Z-track source GX\th 340+0 (Church et al. 2006), and found that it fitted the spectra well 
at all positions on the Z-track, and also clearly suggested a physical model for the phenomenon. In 
this model, the soft apex between NB and FB is a state of the source in which the intensity and  the
mass accretion rate were low. Moving onto the NB, there was a large increase in X-ray intensity 
(count s$^{-1}$) and an increase in the total broadband luminosity by about a factor of two caused by 
an increase in the luminosity $L_{\rm ADC}$ of the Comptonized emission of the ADC. It was argued that 
this increase indicated that $\dot M$ was increasing on the NB and was supported by a marked increase 
in the blackbody temperature of the neutron star from $\sim$1 keV to $\sim$2 keV and even higher values 
on the HB. This heating of the neutron star means a large increase in $T^4$ and so the radiation
pressure became very high. There was a simultaneous decrease in emitting area on the neutron star.
In this situation, it is appropriate to compare the flux emitted
per unit area of the star with the Eddington value (see Sect. 4.1), and the flux increased from a low value 
till on the HB, it was three times super-Eddington. It was argued that this would disrupt
the inner disk deflecting the accretion flow into the vertical direction leading to jet formation, and
explaining why jets are detected on the upper NB and HB where the radiation pressure was high. In flaring 
it was found that $L_{\rm ADC}$ was constant implying that $\dot M$ was constant although the neutron
star luminosity increased. The mass accretion rate per unit area of the neutron star closely agreed with the
theoretical value for the onset of unstable nuclear burning, and it was proposed thus flaring was due to
this. 

In the present work, we apply the extended ADC model to high-quality {\it Rossi-XTE} observations of 
the Z-track source GX\th 5-1 to test the hypothesis that the same physical explanation can explain a 
second Z-track source. Spectral and timing studies were made using the same selections of data allowing
the evolution of kHz QPO frequency to be directly correlated with the spectral evolution. This revealed
a striking dependence of kHz QPO frequency on the neutron star radiation pressure. In the next section
kHz QPO are briefly reviewed.

\subsection{The kHz QPO}

Quasi periodic oscillations (QPO) have been the subject of intense study for many years, and a major 
discovery of the {\it Rossi X-ray Timing Explorer} was that of the twin kilohertz QPO  in LMXB. 
Predictions of millisecond variability pre-date this considerably and Sunyaev (1973) argued that 
hot clumps of material orbiting in the inner disk around a black hole or neutron star would lead to 
quasiperiodic variability on millisecond timescales. Kilohertz QPO have now been detected in many LMXB 
(see reviews of van der Klis 2000, 2006), most of these displaying two peaks at frequencies such as at 
600 and 900 Hz, and in the Z-track sources both frequencies increase during movement along the HB towards 
the hard apex. Rather similar QPO have also been seen in a number of black hole binaries. 

In almost all QPO models the higher frequency peak is associated with the orbital frequency at some 
distance into the accretion disk away from the neutron star; for example, a frequency of 900 Hz corresponds 
to a radial position $r$ of about 18 km. In the sonic point beat frequency model (Miller et al. 1998) 
radiation drag reduces the angular momentum in the disk within a few stellar radii of the neutron star
leading to a sonic point where the radial velocity becomes high (Lamb 1989; Miller 1991; 
Miller \& Lamb 1993, 1996). The upper frequency $\nu_2$ is associated with inhomogeneities orbiting 
at the sonic point in the inner disk, which are irradiated by X-rays from hot regions rotating on the neutron 
star leading to a beating of the upper frequency with the spin frequency $\nu_{\rm spin}$ so producing the 
lower frequency QPO. Although earlier observations found the frequency difference $\Delta \nu$ constant
within errors consistent with the model, improved data showed that $\Delta \nu$ varied around the Z-track. 
The detection of twin kHz QPO and X-ray burst oscillations in the accreting millisecond pulsar SAX\th J1808.4-3658 
(Wijnands et al. 2003) led to the conclusion that $\Delta \nu$ $\sim$ $\nu_{\rm spin}$ in some sources, but 
$\nu_{\rm spin}$/2 in other sources, inconsistent with the model, leading Lamb \& Miller (2003) to 
propose a relativistic disk-spin resonance model.

The relativistic precession model (Stella \& Vietri 1998) identifies $\nu_2$ with motion of inhomogeneities
at the inner disk edge and the lower frequency $\nu _1$ with precession of the inhomogeneities.
The disk resonance model (Abramowicz \& Klu\'zniak 2001) recognizes
the preference of the kHz QPO to have a ratio of 3:2 in their frequencies suggesting
a non-linear resonance between normal modes of the disk. This model predicted that twin peaks
should also exist in black hole binaries, and the discovery of a second peak in
GRO\thinspace J1655-40 (Strohmayer 2001) was a success of the model. Comparison with data
(Abramowicz et al. 2003; Klu\'zniak et al. 2007) for both neutron star and black hole systems
supports the model: black
hole systems accurately follow the 3:2 ratio; LMXB show some scatter but a histogram of the
ratio peaks strongly at 3:2. However, Belloni et al. (2005) have argued that in the case of
\hbox{Sco\th X-1,} the peak in the distribution at a ratio of 1.5 was only marginally significant.

\subsection{The assumed monotonic variation of mass accretion rate}

The assumption of $\dot M$ increasing in the direction HB - NB - FB generally, but not universally made,
is not supported by our previous results for GX\th 340+0, and so we briefly review evidence for and against
this. 

The evidence for this was based originally on a multi-wavelength campaign on Cyg\th X-2 (Hasinger et al. 1990) 
involving X-rays, ultraviolet, optical and radio data. Results from {\it IUE} (Vrtilek et al. 1990) had 
increasing intensity for Z-track movement HB - NB - FB implying an increase of $\dot M$; however, strong 
variability interpreted as flaring may have been due to X-ray dipping in which case there is no strong 
correlation of UV with track position.

The properties of most power spectral features correlate well with position on the Z-track; for example,
the frequencies of the HBO (horizontal branch oscillations), i.e. low frequency QPO, increase around the Z-track 
in the direction HB - NB from $\sim$5 to $\sim$60 Hz (e.g. Homan et al. 2002). Similarly, NBO (normal branch
oscillations) are seen on the NB and in two sources, Sco\th X-1 and GX\th 17+2, oscillations are seen in the FB, 
but only close to the soft apex, and there is a continuous increase of QPO frequency from lower NB to FB 
(Casella et al. 2006), i.e. in crossing the soft apex between these branches. Similarly, the frequencies of 
kilohertz QPO increase along the HB where they are observed (e.g. van der Klis 2000). It is general practice 
to display QPO properties as a function of arc distance $S$ measured along the Z-track in the direction 
HB - NB - FB, and this shows that QPO properties are generally well-correlated with $S$ (e.g. Dieters 
\& van der Klis 2000) implying a dependence on a single parameter, i.e. $\dot M$. It is also generally 
though that the soft apex represents the position at which $\dot M$ increases to the Eddington value 
(Lamb 1989). In addition, the frequencies of HBO and kHz QPO in many models are directly related to 
$\dot M$ (see Homan et al. 2002).

However, in recent years, some authors are beginning to doubt the general assumption of a monotonic
increase in $\dot M$. Some results conflict with the assumption: e.g. in GX\th 17+2, the frequency of the HBO 
first increases, but then {\it decreases} on the normal branch (Wijnands et al. 1996; Homan et al. 2002). 
In Sco\th X-1 in the neighbourhood of the soft apex there is a rapid increase of QPO frequency (Casella et al. 2000; 
Dieters \& van der Klis 2000) with $S$ which is inconsistent with the assumption that $S$ is a measure of 
$\dot M$ as the change is over a very short part of the Z-track. Moreover the approximate constancy
of NBO frequency in general over most of the NB does not indicate a dependency on $\dot M$. Also, in GX\th 17+2, 
the properties of the X-ray bursts sometimes observed do not depend on Z-track position (Kuulkers et al. 2002). 

In our previous work on GX\th 340+0 (Church et al. 2006) we proposed that $\dot M$ increased 
between the soft apex and the hard apex, while on the FB, $\dot M$ was constant. The X-ray intensity,
the measured broadband luminosity of the source and the luminosity of the Comptonized emission
all increased substantially on the NB strongly suggesting that $\dot M$ is not decreasing.
The argument is often made that the X-ray intensity cannot be a good indicator of
$S$ and thus of $\dot M$ (e.g. Dieters \& van der Klis 2000) because the intensity increases on both
the NB and FB, moving away from the soft apex. However, our work on GX\th 340+0 shows this 
not to be a secure argument, since the intensity increases on the NB because $\dot M$ 
increases, while on the FB, $\dot M$ is constant, but strong nuclear burning on the 
neutron star causes the intensity increase. One aim of the present work is to test whether 
similar results are found in GX\th 5-1. 

\subsection{GX\th 5-1}

GX\th 5-1 is the second brightest persistent LMXB (after Sco\th X-1), discovered in 1968 
(Fisher et al. 1968; Bradt et al. 1968) as a bright X-ray source in the Sagittarius region 
and classified as a Z-track source by Hasinger \& van der Klis (1989).  A three-year study of the 
source was made using {\it Ginga} (van der Klis et al. 1991) which revealed an extended horizontal branch 
and an upcurving at the end suggesting a change of behaviour, effectively a fourth branch (Lewin et al. 1992).
Analysis of {\it Exosat} and {\it Ginga} data revealed flaring branch data 
\begin{figure*}[!t]
\begin{center}                                                         
\includegraphics[width=64mm,height=140mm,angle=270]{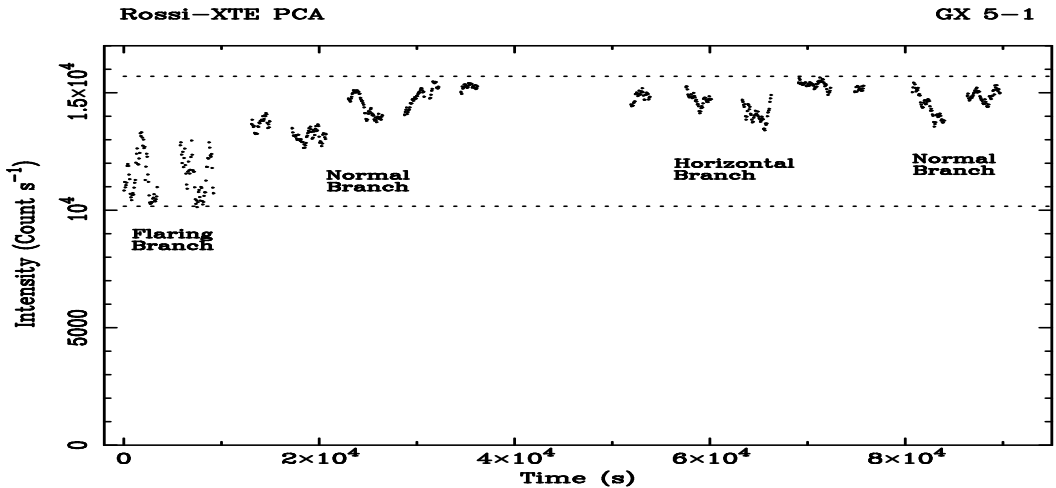}        
\caption{Background-subtracted deadtime-corrected PCA lightcurve of the November 1998
observation of GX\th 5-1 with 64 s binning.}
\end{center}
\end{figure*}
for the first time (Kuulkers et al. 1994) and a detailed
investigation of correlated spectral and timing behaviour was made.  It was the first Z-track source found 
to have QPO (van der Klis et al. 1985) and  kilohertz QPO were detected by Wijnands et al. (1998).
Radio emission has been detected from GX\th 5-1, e.g. by Penninx (1989) implying the existence of jets,
and it is known to be strongest on the horizontal branch (Berendsen et al. 2000), as for 
other Z-track sources. Analysis of 1987 {\it Ginga} data (Asai et al. 1994)
and {\it ASCA} data (Asai et al. 2000) showed there was no evidence for an iron K line, but
a Gaussian feature existed at 10 keV. To date no satisfactory explanation exists for this line.

\section{Observations and analysis}

GX\th 5-1 has been observed several times with {\it RXTE}. We examined the data in the {\sc heasarc}
archive remotely producing hardness-intensity plots and selected data which spanned the full Z-track
in a relatively short time. It is important not to use data in which sideways movement of the Z-track
takes place, broadening the track or even giving several shifted Z-tracks, as we are interested
in the changes taking place around a single Z-track. Similarly data separated by months should not 
be co-added on the assumption that the source was in the same state because portions of the Z-track 
approximately overlap. Thus a single observation including all branches is preferred, and in view of the 
very high count rates in the {\it RXTE} instruments, there is no need to add further data as the 
count statistics are already excellent. We thus selected  the observation of 1998 November 21 - 22 having 
the best coverage of the Z-track in a single observation, consisting of four sub-observations spanning 90 ksec.

Data from both the 
Proportional Counter Array (PCA, Jahoda et al. 1996) and the high-energy X-ray timing experiment (HEXTE) 
were used. The PCA was in Standard2 mode with 16s resolution and examination of the housekeeping data 
revealed that all five of the xenon proportional counter units were on during the observation. 
Standard screening criteria based on the housekeeping data were applied to select data having an offset 
between the source and telescope pointing of less than 0.02$^o$ and elevation above the Earth's limb 
greater than 10$\degmark$. Data were extracted from the top layer of the detector using both left and right 
anodes. Analysis was carried out using the {\sc ftools 5.3.1} package. Lightcurves were generated in the 
band 1.9 - 18.5 keV and background files for each PCA data file produced using the facility {\sc pcabackest}, 
applying the latest ``bright'' background model for Epoch 3 of the mission (1996 April 15 -- 1999 March 22) 
which includes the present observation. Deadtime correction was carried out on both source and 
background files prior to subtraction. Light curves were also made in sub-bands of the above range,
and a hardness ratio defined as the ratio of the intensity in the band 7.3 -- 18.5 keV to that in the 
4.1 -- 7.3 keV band. Tests were made on suitable binning for the lightcurves and Z-track. In order to reduce
scatter in the plot of hardness versus intensity and produce a relatively narrow, well-defined Z-track,
it was found that a binning of 176 s was optimum. In Fig. 1 we show the background-subtracted, 
deadtime-corrected lightcurve in the band 1.9 -- 18.5 keV with the binning reduced from 176 to 64 s to allow
more detail to be seen. In Fig. 2 we show the background-subtracted, deadtime corrected Z-track with 176 s
binning. We identified each subsection of the lightcurve with its position on the Z-track by making 
a hardness-intensity plot for each and finding where this lay on the overall Z-track.
Each of the 14 subsections was selected using the {\sc ftool} {\sc maketime}. It can be seen that
the intensity of the source varied between two well-defined limits corresponding to the Soft Apex
and the Hard Apex of the Z-track as discussed below.

HEXTE data were also extracted as lightcurves 
and spectra from  Cluster 1 of the HEXTE instrument using the {\sc ftool} {\sc hxtlcurve} which also 
provided background files. Deadtime correction was applied using the deadtime coefficients file of 
February 2000. This allowed simultaneous fitting of PCA and HEXTE spectra, the wider energy band 
resulting in spectral fitting being better constrained that with the PCA alone.

\begin{figure}[!h]
\begin{center}                                            
\includegraphics[width=84mm,height=84mm,angle=270]{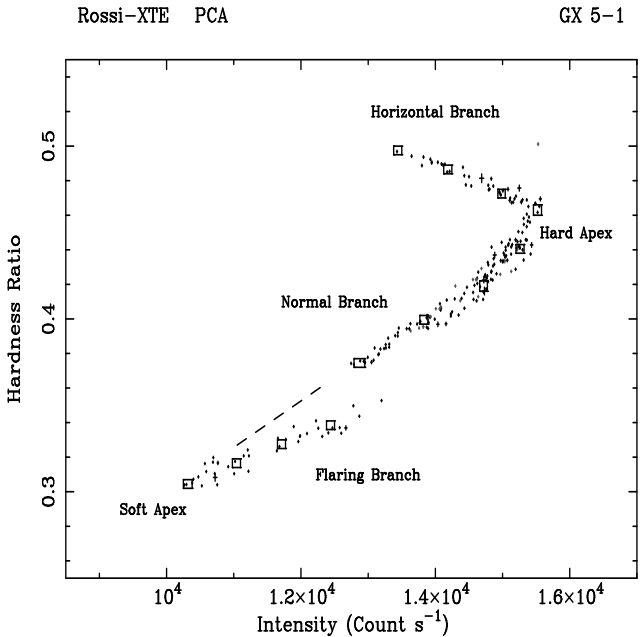}    
\caption{Z-track of the observation derived from background-subtracted and deadtime-corrected lightcurves with
176 s binning. The boxes show the ranges of intensity and hardness ratio used for the selection
of PCA and HEXTE spectra, and timing data.}
\label{}
\end{center}
\end{figure}

\section{Results}

From Fig. 1 the strong variability of GX\th 5-1 during these observations can be seen, and the
lightcurve is labelled to show the identifications with the branches of the Z-track.
At the start of the observation there is strong flaring with individual flares lasting several
thousand seconds. The source then moves to the normal branch with the intensity increasing. 
Following movement along the normal branch to the hard apex, the source reaches the
horizontal branch, the intensity then falling on this branch to intermediate values. 

Figure 2 shows the corresponding Z-track having full coverage of the Z except for the lower part 
of the normal branch, as the source was in this position during a data gap between sub-observations. 
However, this gap was not a problem as it can be seen from the results in Figs. 4, 5 and 6 that 
there is no ambiguity in the behaviour of the source in this part of the Z-track. In the present work,
we carry out both spectral fitting and timing analysis for data selected at a sequence of positions 
along the Z-track. This approach thus allows direct comparison of the QPO results with spectral fitting 
results for exactly the same data.

Our previous work has suggested the importance of selecting data lying on a smooth curve along
the centre of the Z-track (Barnard et al. 2003) so that spectral changes reflect changes along the
track and not perpendicular to it. PCA spectra were generated about equally-spaced along the
Z-track selecting data within boxes in hardness-intensity by use of a good time interval (GTI) 
file for each selection. The boxes were 100 count s$^{-1}$ wide in intensity 
and with a range of hardness ratio of 0.005.
It was checked that the selections used were correct by overlaying the selected data on the full Z-track. 
In Fig. 2 it can be seen that the boxes are small in comparison with the Z-track as a whole
so spectral fitting results relate accurately to a small part of the Z-track. It would clearly 
be inappropriate to use large boxes extending along a significant fraction of the Z-track as
appreciable spectral changes will take place within the box. However, although the boxes used in 
the present work are small, the count rates in the PCA and HEXTE are so large that the total count 
accumulated in each box is very high as shown in Table 1 allowing high quality spectral fitting. 
Accurate determination of QPO frequencies was also possible, this requiring good count statistics as the
fractional modulation is only a few percent.

\tabcolsep 3.5 mm
\begin{table}
\begin{center}
\caption{Counts accumulated in the PCA and HEXTE for each selection; the selections are numbered 1 to 11
starting from the HB end of the Z-track}
\begin{minipage}{80mm}
\begin{tabular}{lrr}
\hline \hline \\
$\;\;$selection& PCA Count& HEXTE Count\\
\noalign{\smallskip\hrule\smallskip}

1   &  2.04$\times 10^6$  &  2.98$\times 10^3$ \\
2   &  4.30$\times 10^6$  &  6.38$\times 10^3$\\
3   &  10.23$\times 10^6$  & 11.69$\times 10^3$ \\
4   &  4.69$\times 10^6$  &  6.66$\times 10^3$ \\ 
5   &  15.71$\times 10^6$   & 15.16$\times 10^3$ \\
6   &  2.17$\times 10^6$  &  3.81$\times 10^3$ \\
7   &  9.78$\times 10^6$  & 10.30$\times 10^3$ \\
8   &  6.78$\times 10^6$  &  6.17$\times 10^3$ \\
9   &  3.05$\times 10^6$  &  4.52$\times 10^3$ \\
10  &  1.63$\times 10^6$  &  2.08$\times 10^3$ \\
11  &  1.74$\times 10^6$  &  1.89$\times 10^3$ \\

\noalign{\smallskip}\hline
\end{tabular}\\
\end{minipage}
\end{center}
\end{table}

After initial tests, it was decided to use 11 spectra consisting of 
three spectra on the horizontal branch, one at the hard apex, four on the normal branch, one at 
the soft apex, and two in flaring, 
\begin{figure*}[!ht]
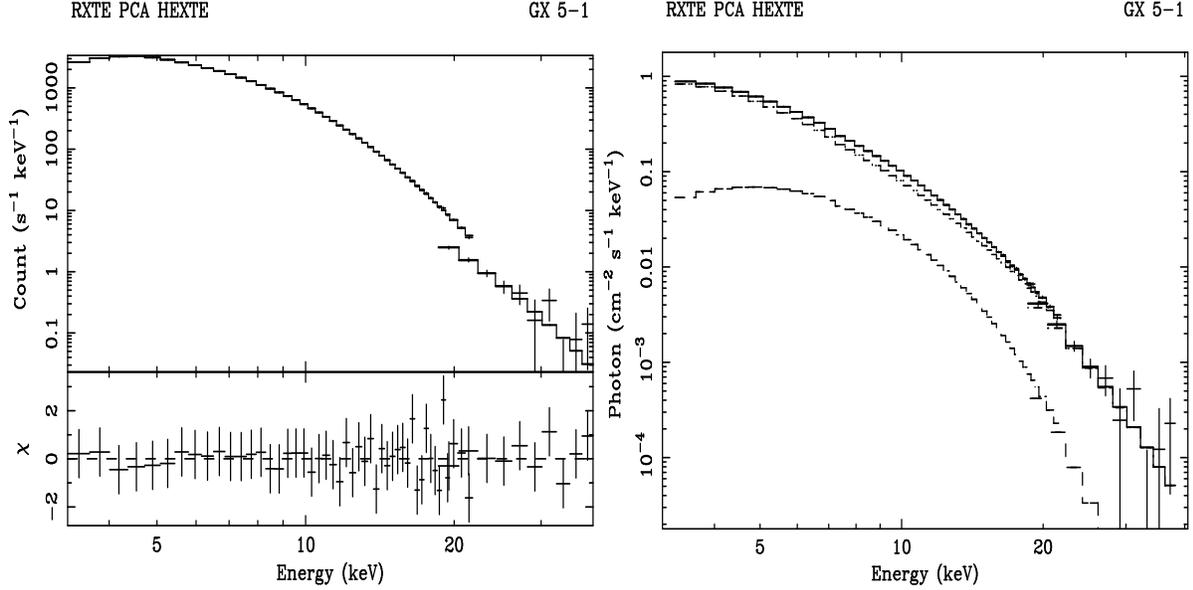
                                                
\begin{center}
\includegraphics[width=78mm,height=78mm,angle=270]{f3a}        
\includegraphics[width=78mm,height=78mm,angle=270]{f3b}
\caption{Left: best-fit to the spectrum of the hard apex with residuals, right: the
corresponding unfolded spectrum}
\label{}
\end{center}
\end{figure*}
numbered as \hbox{spectrum 1} (end of HB) to 11 (FB). 
An extra spectrum close to the peak of flaring (shown in Fig. 2) was found to have
insufficient counts for good spectral fitting. Background spectra were generated for each selection.
Both source and background data were deadtime corrected using a local facility {\sc pcadead}.
Pulse pileup correction has negligible effect on the spectra and so was not made. It was not 
necessary to regroup spectral channels to a specified minimum count to allow use of the $\chi^2$
statistic as the count in all channels was already high. A systematic error of 1\% was added to 
each channel as usual for PCA data. 

HEXTE spectra and background files were produced for each selection using the same GTI files 
and deadtime-corrected using the {\sc ftool} {\sc hxtlcurv}. The auxiliary response file (arf)
of May 2000, and the response matrix file (rmf) of March 1997 were used in spectral analysis.
The rmf file was rebinned to match the actual number of HEXTE channels using {\sc rddescr}
and {\sc rbnrmf}. For both the PCA and HEXTE, the source + background spectra were individually 
compared with background spectra and spectral fitting only carried out up to the energy at which 
these became equal, typically 30 keV in the PCA and 40 keV in HEXTE

\subsection{Spectral analysis}

The motivation of this work is to test the hypothesis that the extended ADC model (Sect. 1)
can provide good fits to the spectra of the Z-track sources, and to test the physical model previously 
proposed to explain the Z-track source GX\th 340+0 (Church et al. 2006) in the case of GX\th 5-1. 
Consequently, the extended ADC  model was used in {\it Xspec} spectral fitting software in the 
form {\sc bb + cpl} as discussed in Sect. 1.

Preliminary fitting showed that the model provided good values of \chisq/d.o.f for all spectra.
The residuals exhibited no evidence for an iron line; however, at the peak of flaring, there was 
evidence for the 10 keV emission feature seen by Asai et al. (1994). This was seen clearly in the 
spectrum eventually discarded because of insufficient counts, but was also detectable in spectra 10 
and 11 on the flaring branch at a significance of about 2$\sigma$. When added to the fitting, the 
line had little effect on the results, for example, changing the neutron star blackbody
radius by about 5\%. Thus the model finally used in fitting all of the spectra omitted the line.

In the Z-track sources, it is well-known that the Comptonized emission has a relatively low cut-off 
energy of a few keV whereas lower luminosity sources have a higher cut-off energy which can approach 100 keV.
Broadband data such as from {\it BeppoSAX} or {\it Suzaku} allow measurement of the cut-off energy and 
the power law photon index $\Gamma$. But in the Z-track sources including GX\th 5-1, the low cut-off 
energy restricts the energy range available for determination of $\Gamma $ and we have taken the approach 
previously adopted (e.g. Church et al. 2006) of fixing $\Gamma $ at 1.7, a physically reasonable value 
(Shapiro et al. 1976). This procedure gave good quality spectral fits, and when a solution was obtained, 
it was found that $\Gamma $ could be freed, but then remained close to 1.7. Extensive testing also revealed that the
results obtained, i.e. the pattern of variation of parameters along the Z-track did not depend
on the value of power law index.

\tabcolsep 3.5 mm
\begin{table*}
\begin{center}
\caption{Spectral fitting results: 90\% confidence errors are shown.}
\begin{minipage}{160mm}
\begin{tabular}{lrrrrlrrr}
\hline \hline
$\;\;$spectrum&$N_{\rm H}$&$kT$&$norm_{\rm BB}$&$R_{\rm BB}$&$E_{\rm
CO}$&$norm_{\rm CPL}$&$\chi^2$/d.o.f.\\
&&keV&&km&keV\\
\noalign{\smallskip\hrule\smallskip}
Horizontal Branch\\
1&6.1$\pm 0.5$  &2.08$\pm 0.10$  &7.6$\pm 0.7$ &5.1$\pm 0.5$         &5.9$\pm 0.3$  &19.5$\pm 1.4$  &32/58\\
2&6.3$\pm 0.5$  &1.98$\pm 0.10$  &7.2$\pm 0.8$ &5.4$\pm 0.6$         &5.8$\pm 0.2$  &21.2$\pm 1.5$  &24/58\\
3&6.5$\pm 0.4$  &1.93$\pm 0.08$  &7.4$\pm 0.8$ &5.8$\pm 0.5$         &5.6$\pm 0.2$  &23.4$\pm 1.5$  &46/58\\
\noalign{\smallskip}
Hard Apex\\
4&6.8$\pm 0.5$  &1.87$\pm 0.09$  &7.2$\pm 0.1$ &6.07$\pm 0.6$        &5.6$\pm 0.2$  &25.3$\pm 1.8$  &38/58\\
\noalign{\smallskip}
Normal Branch \\
5&6.6$\pm 0.4$  &1.77$\pm 0.07$  &6.6$\pm 1.1$ &6.5$\pm 0.5$         &5.4$\pm 0.1$  &25.5$\pm 1.7$  &37/58\\
6&6.5$\pm 0.5$  &1.70$\pm 0.10$  &5.6$\pm 0.6$ &6.5$\pm 0.8$         &5.1$\pm 0.1$  &26.9$\pm 0.8$  &41/58\\
7&6.4$\pm 0.5$  &1.54$\pm 0.07$  &5.0$\pm 0.1$ &7.5$\pm 0.7$         &4.9$\pm 0.1$  &26.1$\pm 2.0$  &40/58\\
8&6.2$\pm 0.5$  &1.41$\pm 0.06$  &4.8$\pm 0.2$ &8.9$\pm 0.8$         &4.7$\pm 0.1$  &24.4$\pm 2.3$  &38/58\\
\noalign{\smallskip}
Soft Apex \\
9&5.7$\pm 0.7$  &1.31$\pm 0.05$  &6.0$\pm 0.2$ &11.3$^{+1.4}_{-0.7}$ &3.7$\pm 0.1$  &25.1$\pm 3.3$  &74/58\\
\noalign{\smallskip}
Flaring Branch\\
10&5.8$\pm 0.7$ &1.39$\pm 0.06$ &8.6$\pm 0.2$ &12.0$\pm 1.0$         &3.6$\pm 0.2$  &25.6$\pm 3.8$  &48/58\\
11&5.9$\pm 0.7$ &1.46$\pm 0.05$ &11.6$\pm 0.1$ &12.7$\pm 0.8$        &3.5$\pm 0.2$  &27.2$\pm 4.0$  &40/58\\
\noalign{\smallskip}\hline
\end{tabular}\\
Column densities are in units of 10$^{22}$ atom cm$^{-2}$; the normalization of the blackbody is in units
of $\rm {10^{37}}$ erg s$^{-1}$ for a distance of 10 kpc, the normalization of the cut-off power law is
in units of photon cm$^{-2}$ s$^{-1}$ keV$^{-1}$ at 1 keV.
\end{minipage}
\end{center}
\end{table*}

The lower energy limit of the PCA, usually set in spectral fitting at $\sim$3 keV, does not
allow very accurate determination of the column density, and it is often the case that $N_{\rm H}$
has to be fixed at the Galactic value. In the present case, the Galactic value of 
$\sim 0.94 \times 10^{22}$ atom cm$^{-2}$ (Dickey \& Lockman 1990) was tried, but the fits
were unacceptable with, for example, \chisq/d.o.f of 288/59 for \hbox {spectrum 1.} Christian \& Swank
(1997) also found a value higher than Galactic ($2.54 \times 10^{22}$ atom cm$^{-2}$). 
In the present case, with free $N_{\rm H}$, the values varied between 
5.7 and 6.8 $\times 10^{22}$ atom cm$^{-2}$ along the Z-track and \chisq/d.o.f. 
became acceptable. Moreover, it was apparent that $N_{\rm H}$ changed systematically
around the Z-track showing not only is the extra absorption intrinsic, but providing evidence
for mass release or outflow within the system associated with the physical changes taking place.
The same effect was seen in the case of {\it RXTE} observations of the Z-track source
GX\th 340+0 (Church et al. 2006). The best-fit to the spectrum of the hard apex is shown in Fig. 3 
as both the folded data (with residuals) and as unfolded data. Final fitting results are shown in Table 1 
and in the following we discuss the behaviour of the blackbody and Comptonized emission.

Firstly, we show in Fig. 4 the variation of the blackbody temperature $kT_{\rm BB}$ and blackbody radius 
$R_{\rm BB}$ around  the Z-track as a function of the total 1 -- 30 keV luminosity $L_{\rm Tot}$
for which a source distance of 9 kpc (Christian \& Swank 1997) was assumed. 
\begin{figure}                                                
\begin{center}
\includegraphics[width=84mm,height=84mm,angle=270]{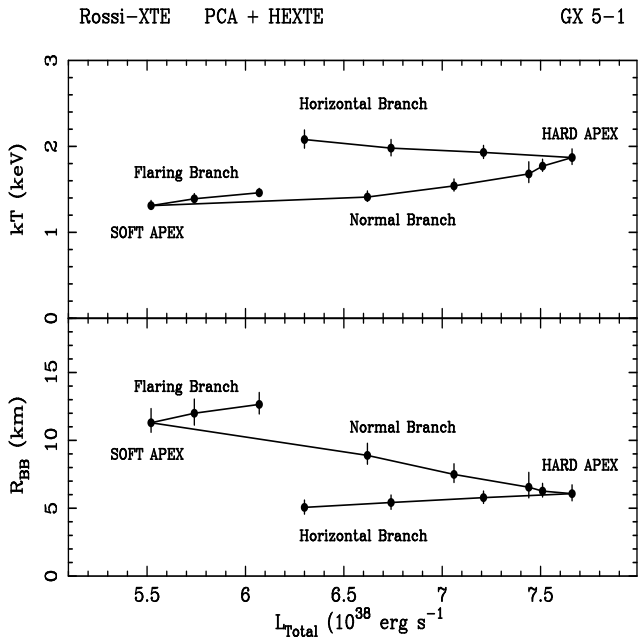}        
\caption{The neutron star blackbody emission: upper panel: the blackbody temperature; lower panel:
the blackbody radius}
\label{}
\end{center}
\end{figure}
The blackbody radius is defined by $L_{\rm BB}$ = 4$\pi \, R_{\rm BB}^2 \sigma T^4$
where $L_{\rm BB}$ is  the blackbody luminosity and $\sigma$ is Stefan's constant. The
blackbody parameters change systematically along the Z, $kT_{\rm BB}$ having its lowest value 
of 1.3 keV at the soft apex where $R_{\rm BB}$ is maximum at 11$^{+1.4}_{-0.7}$ km.
Moving along the normal branch, $kT_{\rm BB}$ increases to 1.9 keV at the hard apex, and 
continues to increase on the horizontal branch, reaching 2.1 keV at the end of the Z-track. 
At the soft apex, the value of $R_{\rm BB}$ indicates that the whole neutron star is
emitting, but decreases on the normal branch showing that the emission contracts presumably to
an equatorial belt. In Sect. 4.1 we discuss this measured decrease in emitting area on the neutron star 
showing that it is consistent with the effects of increased radiation pressure.
\begin{figure}[!ht]                                                
\begin{center}
\includegraphics[width=84mm,height=84mm,angle=270]{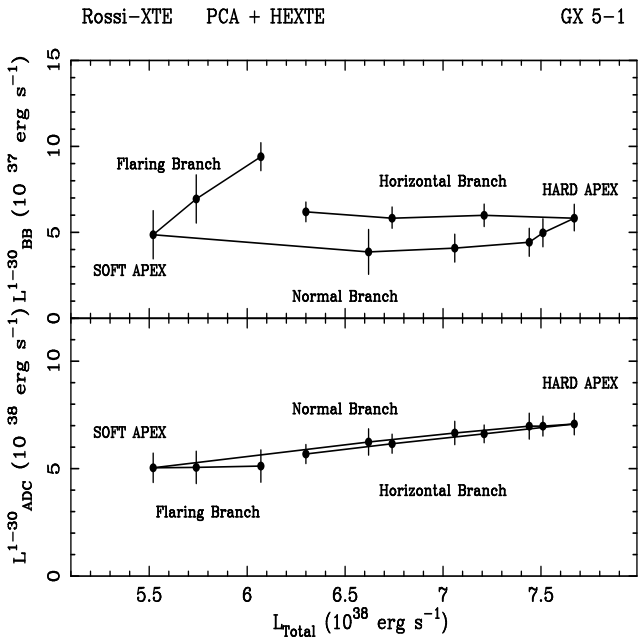}        
\caption{Variation of emission component 1 -- 30 keV luminosities around the Z-track: upper panel: the blackbody
luminosity; lower panel: the ADC Comptonized emission luminosity}
\label{}
\end{center}
\end{figure}
Thus at the soft apex the source appears to be in an undisturbed state with the neutron star emitting
from its whole surface at its lowest temperature; we see below that the evidence indicates the mass 
accretion rate is minimum at this position.

\tabcolsep 3.5 mm
\begin{table}
\begin{center}
\caption{Total luminosity, blackbody luminosity and ADC luminosity in the band 1 - 30 keV
in units of erg s$^{-1}$.}
\begin{minipage}{80mm}
\begin{tabular}{lrrr}
\hline \hline \\
$\;\;$spectrum  & $L_{\rm Tot}$ & $L_{\rm BB}$   & $L_{\rm ADC}$\\
&($10^{38}$) & ($10^{37}$) & ($10^{38}$)\\

\noalign{\smallskip\hrule\smallskip}

1   & 6.30    & 6.19$\pm$0.57   & 5.68$\pm$0.44   \\
2   & 6.74    & 5.82$\pm$0.61   & 6.16$\pm$0.44   \\
3   & 7.21    & 5.99$\pm$0.65   & 6.62$\pm$0.41   \\
4   & 7.67    & 5.82$\pm$0.77   & 7.08$\pm$0.50   \\
5   & 7.51    & 4.97$\pm$0.81   & 6.98$\pm$0.46   \\
6   & 7.44    & 4.42$\pm$0.81   & 6.98$\pm$0.60   \\
7   & 7.06    & 4.08$\pm$1.30   & 6.66$\pm$0.54   \\
8   & 6.62    & 3.86$\pm$1.40   & 6.24$\pm$0.61   \\
9   & 5.52    & 4.86$\pm$1.40   & 5.04$\pm$0.68   \\
10  & 5.74    & 6.94$\pm$1.40   & 5.06$\pm$0.75   \\
11  & 6.07    & 9.40$\pm$0.81   & 5.12$\pm$0.75   \\

\noalign{\smallskip}\hline
\end{tabular}\\
As is normal practice, we do not quote errors in the total luminosity.
\end{minipage}
\end{center}
\end{table}

In Fig. 5 we show the variation of the luminosities of the individual emission components $L_{\rm BB}$
and $L_{\rm ADC}$: the neutron star blackbody and the Comptonized emission of the ADC. Firstly, it is clear
that the Comptonized emission is dominant, its luminosity being ten times larger than that of the
blackbody. However, in flaring, the blackbody luminosity increases by a factor of two to $\sim$40\% of the total 
at the peak of flaring. Moving away from the soft apex on the normal branch there is a strong increase 
of $L_{\rm ADC}$ by 40\% on the normal branch, with the PCA intensity increasing from $\sim$10\th 000 to 
$\sim$15\th 500 count s$^{-1}$. The increase in X-ray intensity is, of course, unquestionable. 

The significance of the ADC luminosity increase on the normal branch was assessed by fitting the 
variation of $L_{\rm ADC}$ with $L_{\rm Tot}$ (lower panel of Fig. 5) also shown in Table 3 using 
only the normal branch data. 
A linear fit gave a slope of 0.945$\pm$0.35, where the errors at 90\% confidence strongly exclude a slope
of zero. The probability that the slope is zero (i.e. testing the hypothesis that the luminosity is constant)
is less than 0.006\% so that this increase of
ADC luminosity on the normal branch is highly significant.
%
%
Fig. 5 shows that the increase of X-ray intensity in climbing the normal branch is due to the increase in $L_{\rm ADC}$.
We suggest that it is {\it unlikely} that $L_{\rm ADC}$ could increase so substantially without an increase 
in mass accretion rate $\dot M$, and this is also indicated by the increase in neutron star blackbody
temperature. In previous work, it was often assumed that the Z-track was produced by the mass accretion rate 
$\dot M$ varying  monotonically in the direction HB - NB - FB (Sect. 1), although this conflicted 
with the known {\it increase} of X-ray intensity between the soft and hard apex.
We propose here based on the measured definite increase of $L_{\rm ADC}$ that $\dot M$ increases
on the normal branch between the soft apex and hard apex, in the opposite direction to that normally
assumed. The behaviour of the neutron star blackbody is more complicated consisting of a temperature
increase but an area decrease such that the overall luminosity does not change substantially as discussed
in Sect. 4.1 and so we should not expect $L_{\rm BB}$ to follow $\dot M$ in a simple way.

The generally accepted idea that $\dot M$ increases monotonically around the Z-track explains flaring 
as a luminosity increase due to high and presumably variable $\dot M$. However, it is clear from Fig. 5 
that $L_{\rm ADC}$ is constant within errors on the FB indicating that $\dot M$ does not change in flaring.
The blackbody luminosity however, clearly increases and we propose that flaring must consist of 
luminosity increase due to unstable thermonuclear burning on the surface of the neutron star, a
suggestion that has not previously been made. During flaring, there is a relatively small increase of 
blackbody temperature, but the blackbody radius rises to $\sim$13 km for spectrum 11, and would be even larger at
the peak of flaring. Values of $R_{\rm BB}$ substantially larger than the probable neutron star radius have been
found previously in GX\th 5-1 (Christian \& Swank 1997), and a type of radius expansion known in X-ray bursts 
may be taking place, with the emitting region expanding beyond the normal surface of the neutron star.

\begin{figure}                                                
\begin{center}
\includegraphics[width=84mm,height=84mm,angle=270]{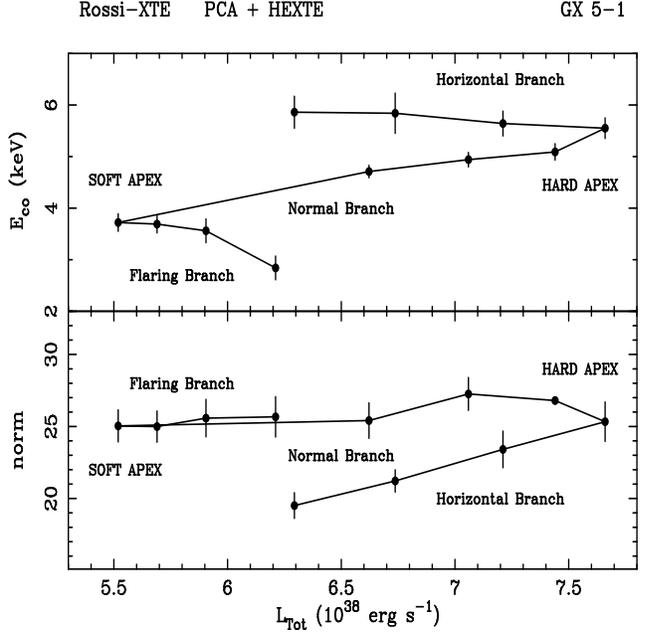}        
\caption{The cut-off energy (upper panel) and normalization (lower panel) of the ADC Comptonized emission.}
\label{}
\end{center}
\end{figure}
Figure 6 shows the variation of the ADC Comptonized emission parameters: the normalization and the cut-off energy 
$E_{\rm CO}$. Firstly, there is an increase of normalization moving on the NB away from the soft apex,
while $E_{\rm CO}$ increases, these factors combining to give the
observed increase of $L_{\rm ADC}$ (Fig. 5). Secondly, comparison of Figs. 4 and 6 shows that 
$E_{\rm CO}$ is well correlated with $kT_{\rm BB}$ on the HB and NB,
suggesting a link between the emission from the neutron star and the electron temperature of the
ADC. For an ADC with high optical depth to electron scattering $E_{\rm CO}$ = $3\,kT_{\rm e}$ 
(Petrucci et al. 2001) whereas for low optical depth $E_{\rm CO}$ = $kT_{\rm e}$. The present results
allow a comparison of the neutron star blackbody temperature and $E_{\rm CO}$ and it is found that
$E_{\rm CO}/kT_{\rm BB}$ has an average value of 2.7$\pm $0.5 implying that the plasma is optically
thick and that the ADC electron temperature is close to the neutron star blackbody temperature.
The relationship of $kT_{\rm BB}$ and $kT_{\rm e}$ in LMXB was discussed by Ba\l uci\'nska-Church \& Church (2005)
who showed that in lower luminosity sources the ADC electron temperature $kT_{\rm e}$ was substantially
higher than $kT_{\rm BB}$ so that there must be a presently-unknown heating process. However, in
brighter sources ($L$ $\approxgt$ $2\times 10^{37}$ erg s$^{-1}$), these two temperatures were found to be
consistently equal, providing evidence for thermal equilibrium between the neutron star and corona.

\begin{figure}[!ht]                                                  
\begin{center}
\includegraphics[width=84mm, height=84mm, angle=270]{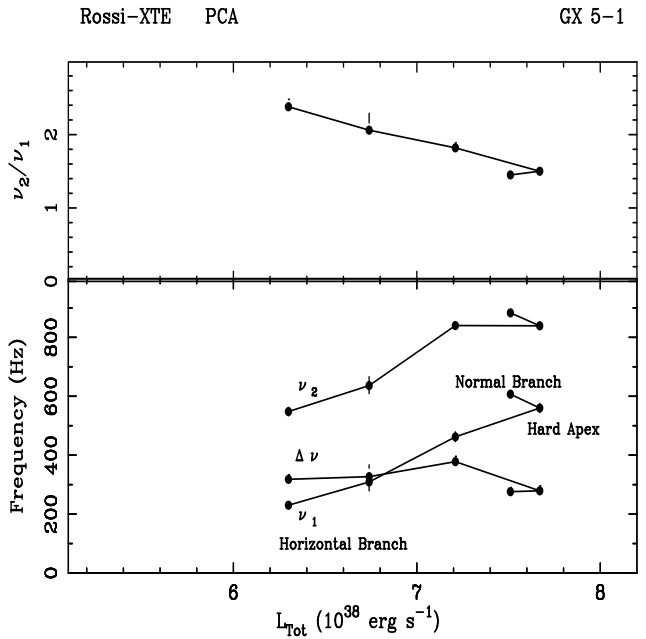}
\caption{Lower panel: The variation of lower frequency and upper frequency kHz QPO detected on the upper normal
and horizontal branches as a function of the total 1 - 30 keV luminosity ($L_{\rm Tot}$) together with
the difference $\Delta \nu$; top panel: variation of the frequency ratio $\nu_2/\nu_1$ around the Z-track}
\end{center}
\end{figure}

\subsection{Timing analysis}

We next examine the results of carrying out timing analysis for the same selections of data along the
Z-track as used in Sect. 3.1. Two types of data had been accumulated suitable
for timing analysis up to kHz frequencies: event mode data of type E\_125us\_64M\_24\_1s
and single bit data of types SB\_125us\_0\_13\_1s, SB\_125us\_14\_17\_1s and SB\_125us\_18\_23\_1s,
i.e. one mode covering the energy band up to 20 keV stored in three files. Power density spectra
were extracted with 125 $\mu$s time resolution from both event mode and single bit data
into a single power spectrum for each selection along the Z-track. The energy range was restricted 
to the band 5 - 60 keV for event mode data with the {\it RXTE}-specific ftool {\sc sebitmask} specifying
channels corresponding to these energies and applying this with {\sc feselect} to the data files.
Power spectra were then extracted using the {\it Xronos} facility {\sc powspec} which essentially
rebinned all data into a single energy bin. A standard Leahy normalization was applied. In order 
to carry out high quality fitting of the power spectra, we converted the spectra
to a format that allowed fitting with the {\it Xspec} spectral fitting software, requiring changes
to the header of the {\it fits} files and re-naming of the columns to correspond to frequency, power and
power error. Finally an output file was made suitable for use with {\it Xspec} using
the ftool {\sc flx2xsp} producing a file with {\it pha} extension as for an energy spectrum. This
ftool also created a dummy instrument response function as needed by {\it Xspec} consisting of a unitary
matrix having no effect on the data. Deadtime correction was carried out and the spectra grouped to
a minimum of 60 counts per bin as found suitable by trial and error. 

The power spectra were fitted within the range 150 - 950 Hz, using a three-component model consisting of
two Lorenzian lines for the kilohertz QPO plus a power law which was able to fit the continuum of the spectrum
well within the restricted frequency range.
As expected, kHz QPO were only detected on the horizontal branch of the Z-track and the upper part of 
the normal branch, that is, from a total of 5 of the 11 positions at which spectral fitting was carried 
out. Fig. 7 (lower panel) shows the best-fit frequencies obtained for the two QPO peaks
in these spectra as a function of total luminosity ($L_{\rm Tot}$) in the band 1- 30 keV obtained
previously during spectral fitting. It can be seen that there is a monotonic systematic increase of
the QPO frequency from the end of the HB to the point on the upper NB 
for both the upper and lower frequency QPO. Thus we find the same general trend of the kHz QPO frequencies
as reported previously for GX\th 5-1 (Jonker et al. 2002) and in other Z-track sources (e.g. Jonker et al.
2000). The frequency difference $\nu_2$ - $\nu_1$ is also shown in the lower panel and can be seen
to have a minimum of $\sim$280 Hz at the hard apex and a maximum of $\sim$380 Hz part way along the horizontal
branch. In terms of the 90\% confidence uncertainties in $\nu_2$ and $\nu_1$ obtained in the fitting, 
typically 10 - 30 Hz, the variation is significant. The ratio of frequencies $\nu_2/\nu_1$ also shown 
in the upper panel.
Although the ratio passes through the value 1.5, i.e. the ratio 3:2 significant in the resonance
model of QPO (Sect. 1), and interestingly this occurs at the hard apex, the ratio varies substantially
around the Z-track from 1.4 on the normal branch to 2.4 at the end of the horizontal branch, only
equalling 1.5 in one part of the track.

\begin{figure}[!ht]                                                  
\begin{center}
\includegraphics[width=84mm, height=84mm, angle=270]{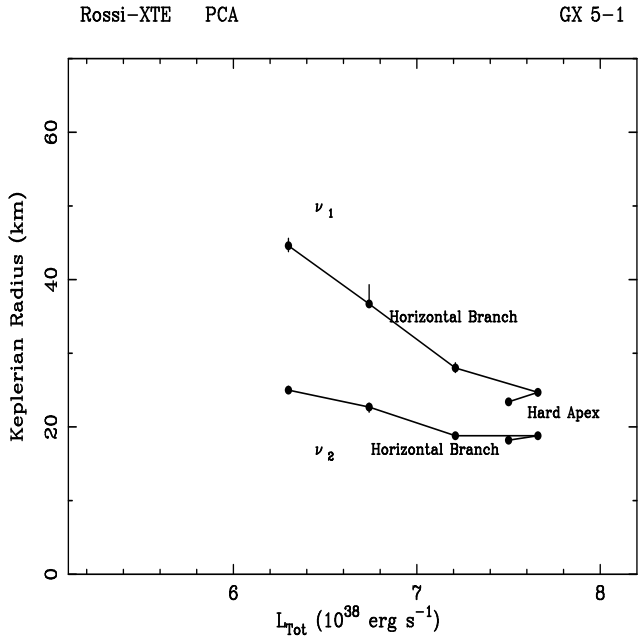}
\caption{The orbital radii of the upper and lower frequency QPO as a function of luminosity}
\end{center}
\end{figure}

We next assume that the upper frequency $\nu_2$ corresponds to an orbital frequency in the inner disk
and calculate the orbital radii for both kHz QPO assuming Newtonian dynamics in which the orbital
frequency at a radius $r$ in the disk is given by

\[2\, \pi \, \nu = \sqrt{{G\, M\over r^3}}.\]

\noindent where $M$ is the mass of the neutron star assumed to be 1.4$\,$M$_{\sun}$. These radial 
positions are shown in Fig. 8. For the higher frequency QPO, the frequency $\nu_2$ for the power 
spectrum in the normal branch corresponds to a radial position of about 18 km, i.e. some distance 
away from the neutron star surface.

As the source moves on to the horizontal branch, the frequency decrease
corresponds to a radial position moving outwards eventually reaching 25 km clearly suggesting that
the source of oscillation moves further away from the neutron star. In terms of the spectral
evolution along the Z-track, this is very suggestive, since both in the present work (Sect. 4.1;
Fig. 9) and in our previous work on GX\th 340+0 (Church et al. 2006), the radiation pressure
at the inner disk becomes very strong on the HB and is expected to disrupt the inner disk
removing the inner portions. Discussion of this will be postponed until after more detailed
consideration of the spectral fitting results.

\section{Discussion}

\subsection{Spectral fitting results}

The results obtained in Sect. 3.1 show that the spectra of GX\th 5-1 are well-fitted by the extended ADC 
model, and that the evolution of spectral parameters around the Z-track suggest a relatively simple
explanation of the Z-track as discussed below. The increase of ADC luminosity between the soft apex and 
hard apex provides evidence that $\dot M$ increases in this direction. The constancy of the ADC luminosity 
on the flaring branch combined with an increasing $L_{\rm BB}$ shows that flaring {\it must} consist of 
unstable nuclear burning on the neutron star. Both of these conclusions disagree with the often-held view 
that $\dot M$ increases monotonically in the direction HB -- NB -- FB. Moreover, the spectral evolution of
GX\th 5-1 around the Z-track is essentially the same as in the similar source GX\th 340+0
(Church et al. 2006) so that the same explanation works for both sources. We discuss below
effects taking place on the normal and horizontal branches and on the flaring branch in more detail.

\subsection{The effects of radiation pressure}

Moving away from the soft apex where the neutron star blackbody temperature is lowest and the blackbody
radius indicates that the whole star is emitting, we find a substantial increase in $kT_{\rm BB}$, so that
there is a large increase in the radiation pressure $\sim $ T$^4$.
\begin{figure}[!ht]
\begin{center}                                                
\includegraphics[width=84mm,height=84mm,angle=270]{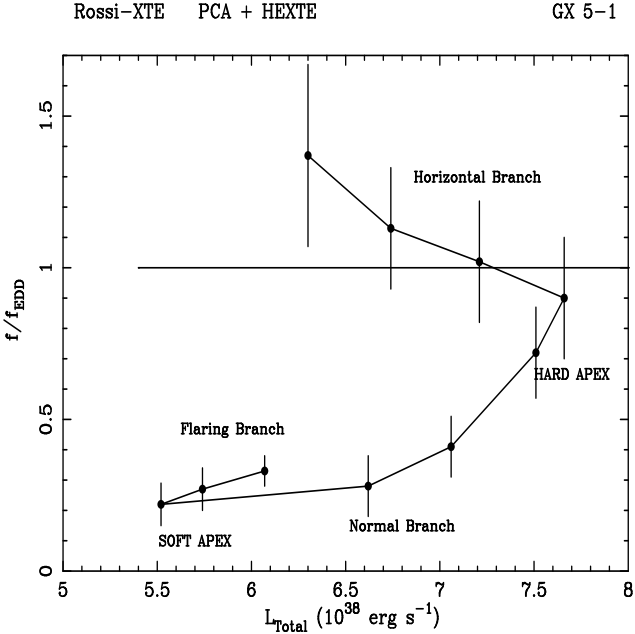}        
\caption{The ratio of neutron star emitted flux to the Eddington flux as a function of position on
the Z-track.}
\end{center}
\end{figure}

At the same time, the blackbody radius $R_{\rm BB}$ decreases to a value of $\sim$5 km at the HB end of 
the Z-track so that the emission is from a reduced area on the neutron star, presumably an equatorial belt. 
It would be normal to describe the source in terms of its luminosity compared with the Eddington luminosity
which is, of course, derived on the assumption of radial and not disk accretion and assuming accretion
over 4$\pi$ steradians. In the present case, where only parts of the neutron star emit, it is instructive 
to compare the flux $f$ emitted by unit emitting area on the neutron star with the Eddington flux $L_{\rm Edd}/4\pi R^2$, 
where $R$ is the neutron star radius assumed to be 10 km, so that $f_{\rm Edd}$ = 
$1.4\times 10^{25}$ erg cm$^{-2}$ s$^{-1}$. When only parts of the neutron star are emitting, the flux $f$
will exceed the mean X-ray flux deduced from the measured luminosity and the source distance. Figure 9 shows the 
ratio of $f$ to $f_{\rm Edd}$. The ratio has its lowest value at the soft apex of $\sim$0.2 showing that radiation 
pressure is weak, rising to unity at the hard apex and continuing to rise to the end of the Z-track. Thus the 
radiation pressure of the emitting band becomes strong at the hard apex and horizontal branch such that 
the effects on the accretion flow will be strong. At the Eddington flux, the radiation pressure force balances 
the inwards gravitational force, both flux and gravitational force decreasing as $1/r^2$ as in the standard 
derivation of the Eddington limit, except that this does not apply over the whole neutron star surface
and the effects of radiation pressure will depend on which direction is considered.

Especially in luminous LMXB, the inner accretion disk is radiatively supported by the radiation pressure
of the disk itself having an equilibrium half-height $H_{\rm eq}$ = $3\sigma _{\rm T} \dot M/8 \pi m_{\rm p}\,c$,
where $\sigma _{\rm T}$
is the Thomson cross section, $m_{\rm p}$ is the proton mass and $c$ the velocity of light (Frank et al. 2002).
The height of the disk rises rapidly from zero to  $H_{\rm eq}$ in a radial distance $\sim$10 km
from the surface of the neutron star, and for the hard apex in the present data, $H_{\rm eq}$ is 65 km
where $\dot M$ is derived from the total 1 -- 30 keV luminosity. Thus the inner disk
rises high above the neutron star, and we propose that the strong radiation pressure at the hard apex
and on the HB severely disrupts the inner disk blowing large parts of it away. At any radial and vertical position 
in the unperturbed inner disk ($r$, $z$), the vertical force due to radiation pressure plus gas pressure
is balanced by the vertically downwards component of the gravitational force. Radiation pressure from the neutron
star acts on a line from the star to this point ($r$, $z$) offsetting the gravitational force. If this offset
reduces the effective gravitational force close to zero, the equilibrium height of the disk would tend to
infinity. Moreover, if we assume that $\dot M$ is maximum
on this part of the Z-track, the radiation pressure of the disk itself will also be increased moving
the disk surface upwards. 

Clearly the combined effects of the radiation pressure of the neutron star and that of the inner 
accretion disk will be complicated and modelling should be carried out. That is beyond the scope
of the present work, but we show in Fig. 10 an approximate schematic view of the expected effect.
The unperturbed profile of the disk is shown as a dashed line, and the actual profile of the inner disk 
by a full line. The disk height $H(r)$ will revert to the unperturbed form of the 
radiatively-supported disk at some radial position $r_{\rm un}$, and this position 
represents an effective inner edge of the disk, since at smaller radial positions there may be a 
residual disk, but of substantially reduced height.

The increase of column density found in spectral fitting on the normal branch supports the release of mass 
within the system on this part of the Z-track. Our measured increase of absorption indicates the absorber
is spatially extended so that both emission components will be affected. It is clear that most
of this material would not be fully ionized since the ionization parameter $\xi$ = $L/n\,r^2$ decreases 
rapidly with distance and a $\xi$ of 10$^6$ is required for there to be complete ionization (e.g. Makishima 
1986). Moreover there would be a spatial variation of $\xi$ so that material located at vertical positions
not directly facing the emitting part of the neutron star would be less ionized.

\begin{figure}[!t]
\begin{center}                                                      
\includegraphics[width=50mm,height=70mm,angle=270]{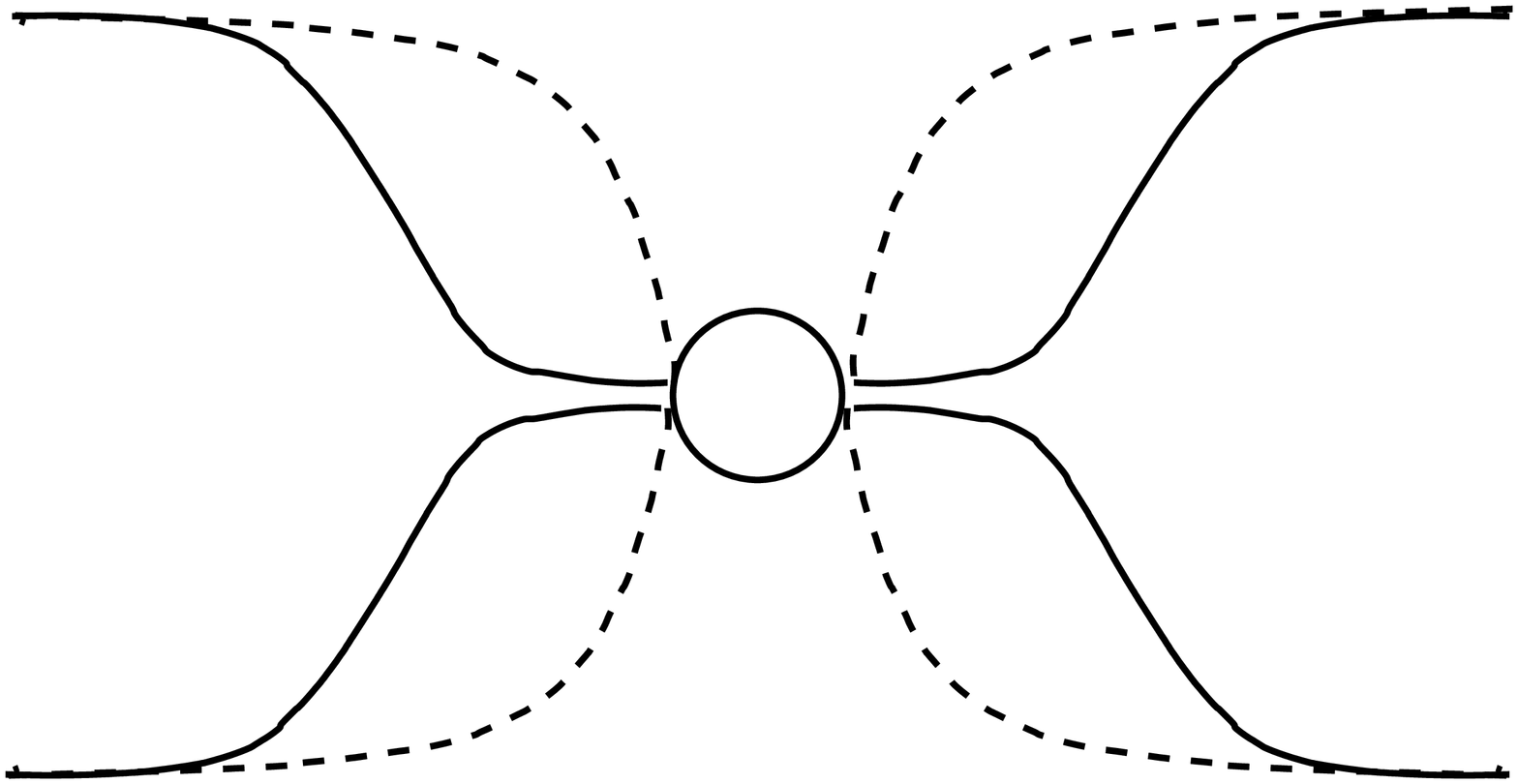}        
\caption{Schematic view of the inner, radiatively supported inner accretion disk
disrupted by radiation pressure; the unperturbed disk is shown as a dashed line.
The radial profile of height $H(r)$ meets the unperturbed profile at radial position $r_{\rm un}$ and 
it is proposed in the present work that the higher frequency kHz QPO consists of an oscillation taking place
in the neighbourhood of $r_{\rm un}$}
\end{center}
\end{figure}

The localized radiation pressure of the neutron star as measured by $f/f_{\rm Edd}$ rises by almost an 
order of magnitude from the soft apex to the end of the HB. Radio emission is non-existent on the FB and 
soft apex and strongest at the hard apex and HB. Thus in GX\th 5-1 as found previously in GX\th 340+0
there is a very good correlation between observation of radio and $f/f_{\rm Edd}$ and we propose
that strong radiation pressure is a necessary, but not necessarily sufficient condition for jet formation.

We also comment on the observed decrease of blackbody radius, i.e. emitting area on the neutron star in 
ascending the normal branch, noting that this is consistent with previous work. The blackbody radius
obtained from spectral fitting $R_{\rm BB}$ is related to the half-height of the emitting equatorial strip on
the neutron star $h$ by $4\, \pi\, R_{\rm BB}^2$ = $4\, \pi\, R_{\rm NS}\, h$, where $R_{\rm NS}$ 
is the radius of the neutron star, assumed to be 10 km. In a survey of LMXB with {\it Asca} and {\it BeppoSAX}, 
Church \& Ba\l uci\'nska-Church (2001) made a survey comparing the blackbody luminosity ($L_{\rm BB}$) in 
the survey sources with the Comptonized luminosity ($L_{\rm ADC}$) in order to investigate the well-known
discrepancy that $L_{\rm BB}$ in most LMXB is many times less than the fraction of the total luminosity
expected ($\sim$50\%) on the basis of simple theory. It was found that the half-height of the 
equatorial emitter on the neutron star $h$ was in the lower luminosity LMXB substantially less than the 
neutron star radius. However, in all sources $h$ followed the simple relation  $h$ = $H$, where $H$ is the 
half-height of the inner disk, valid over 3 decades of source luminosity. A probable explanation 
of this geometric equality is the model of accretion flow spreading on the neutron star (Inogamov \& 
Sunyaev 1999), which was shown to agree reasonably with the survey results (Church et al. 2002). Thus 
in lower luminosity sources, the spreading does not extend so far vertically, and the emitting region 
is smaller.

We have suggested that in the Z-track sources parts of the inner accretion disk are blown away by high 
radiation pressure presumably leaving a residual innermost disk of reduced height. Thus a reduced value of 
disk height $H$ is expected to lead to a reduced emitter size on the neutron star, and conversely, the 
observed decrease of $R_{\rm BB}$ and $h$ is evidence that the height of the inner disk $H$ is reduced 
from more than 50 km to $\sim$3 km indicating disruption of the inner disk.
 

\subsection{Flaring}

The constancy of $L_{\rm ADC}$ on the flaring branch and the increase of $L_{\rm BB}$ lead to the conclusion
that flaring must be unstable nuclear burning on the neutron star, since if $\dot M$ increased as usually 
assumed we would definitely expect $L_{\rm ADC}$ to increase substantially as it does on the normal branch. 
Further support for this
conclusion comes from comparing with the theory of unstable nuclear burning. Unstable burning is expected at
various mass accretion rates (Fujimoto et al. 1981; Fushiki \& Lamb 1987; Bildsten 1998; Schatz et al. 1999),
the various r\'egimes of stable and unstable burning depending on the value of $\dot m$, the mass accretion
rate per unit area of the neutron star. For $5.0\times 10^3 < \dot m < 1.3\times 10^5$ g cm$^{-2}$ s$^{-1}$
there is unstable He burning in a mixed H/He environment, while for $\dot m > \dot m_{\rm ST}$ =
$1.3\times 10^5$ g cm$^{-2}$ s$^{-1}$, burning is always stable (Bildsten 1998), this value having an estimated
uncertainty of 30\% (giving a range of about 0.9 - 1.7$\times 10^5$ g cm$^{-2}$ s$^{-1}$). In deriving values 
of $\dot m$, we use actual values of the emitting area given by the blackbody radius obtained in spectral fitting,
i.e. $\dot m$ = $\dot M/(4\, \pi\, R_{\rm BB}^2)$. At the soft apex, we find
$\dot m$ = $1.8\pm 0.2\times 10^5$ g cm$^{-2}$ s$^{-1}$ consistent with the critical value
$\dot m_{\rm ST}$, and so it is expected theoretically that burning is stable on the normal branch but unstable
on the flaring branch, the soft apex being the point at which the transition between these occurs.

If we progress along the FB away from the soft apex, the value of $\dot m$ derived from spectral fitting
decreases, so that once started, unstable burning remains unstable. Moving in the opposite direction away
from the soft apex, however, $\dot m$ rises substantially, to $3.6\pm 0.5 \times 10^5$ g cm$^{-2}$ s$^{-1}$
for the first spectrum on the NB, increasing eventually to $10\times 10^5$ g cm$^{-2}$ s$^{-1}$
at the end of the HB. This is due to the decrease of blackbody emitting area and the increase of $\dot M$
and shows that conditions on the neutron star are well into the r\'egime of stable burning.

It would be desirable to find other observational tests of whether flaring consists of nuclear burning
and it would be
interesting to derive an $\alpha$-parameter for flaring in GX\th 5-1, since the agreement of measured
$\alpha$ values with theory was the essential proof that bursting was nuclear in nature. $\alpha$ is defined
as the ratio of accretion luminosity to burst luminosity evaluated in bursting by integrating the non-burst
luminosity between bursts, and the luminosity within a single burst. For a single burst, the time needed for 
the accretion flow to accumulate and to settle on the neutron star increasing in density towards ignition conditions 
is well-defined as the time between bursts. Theoretical values are obtained by dividing the accretion luminosity 
per nucleon ($GM/R_{\rm NS}$ by the nuclear energy release 
per nucleon from which $\alpha$ $\sim$30 for H-burning and more than 100 for He burning. However, a detailed 
investigation of flaring in the Z-track sources in the context of nuclear burning has not been made and the 
physics of the accumulation of mass and settling on the star has not been investigated and it is unclear what 
timescale is involved in the matter involved coming to ignition conditions. Thus consideration of the 
$\alpha$-parameter is beyond the scope of the present work.

\subsection{The kHz QPO}

We are now able to discuss the QPO results in terms of the spectral fitting results, and we firstly 
take the QPO variation along the Z-track shown in Fig. 8 as the Keplerian radius of the oscillation
versus $L_{\rm Tot}$ and re-draw this as Keplerian radius as a function of $f/f_{\rm Edd}$ as shown in
Fig. 11.

A striking feature can be seen that the major change in Keplerian radius and thus in QPO frequency takes
place when $f/f_{\rm Edd}$ equals unity. This result is unlikely to be coincidental and suggests
that the strong radiation pressure causes the variation in QPO frequency along the Z-track.
Moving from the hard apex to the end of the Z-track, $f/f_{\rm Edd}$ increases from 1.0 to 1.5 and the 
Keplerian radius of the higher frequency QPO increases as the frequency decreases. The most obvious 
explanation is that this radiation pressure is indeed disrupting the inner disk so that the oscillations 
comprising the higher frequency QPO move outwards but staying on the inner edge or close to it, and this 
can explain the change of frequency.  It is clear that the lower frequency QPO responds to the change
in $\nu_2$ in some way but the results do not provide a clear indication of the mechanism of this. 
However, the result shown in Fig. 11 does suggest an alternative to existing ideas regarding possible truncation
of the inner disk. Much theoretical work suggests that the accretion disk is in contact with the neutron star,
e.g. the Inogamov \& Sunyaev model (1999) of accretion flow spreading on the surface of the neutron star,
while other work considers the possibility that the disk is truncated at a specific radius, either at an 
innermost stable orbit,
or at an Alfv\'en radius assumed external to the neutron star by capture onto the weak magnetic field of
the neutron star or at a sonic point. In the sonic point model (Miller et al. 1998), part of the
accretion flow can remain in the disk until it impacts on the neutron star. The present work suggests 
that the disk may be truncated, but only
in part of the Z-track, and that the mechanism may be radiation pressure.

\begin{figure}[!ht]                                                  
\begin{center}
\includegraphics[width=80mm, height=80mm, angle=270]{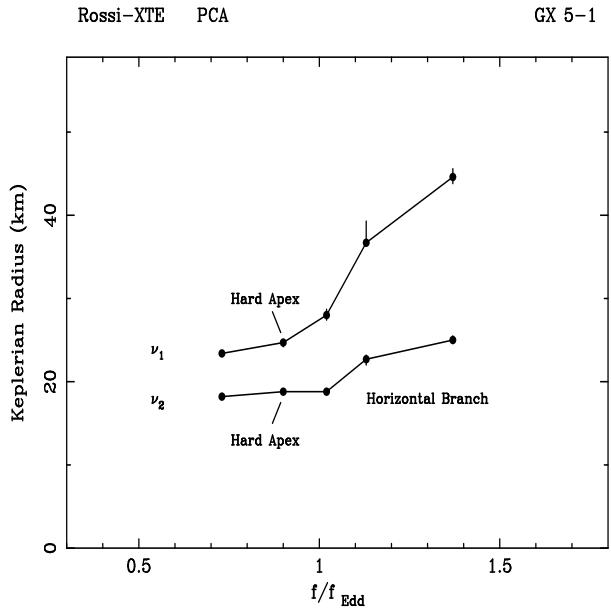}
\caption{Variation of Keplerian radius of the kHz QPOs as a function of radiation pressure of the
neutron star, i.e. the emitted flux of the surface in units of the Eddington flux $f_{\rm Edd}$
demonstrating a striking correlation between the QPO radial position of origin and high radiation pressure}
\end{center}
\end{figure}

\section{Conclusions}

We have shown that spectra selected along the Z-track in GX\th 5-1 are well fitted using the extended 
ADC model, and that the spectral evolution is very similar to that we found in GX\th 340+0 (Church
et al. 2006). Thus the physical model proposed to explain the spectral evolution in that source is
also able to explain GX\th 5-1. In this model, the soft apex has the lowest value of $\dot M$ and the strong
physical changes on the normal branch are primarily due to an increase of $\dot M$ and of neutron star
temperature. The increase of temperature means that the radiation pressure {\it must} be high
in the neighbourhood of the neutron star. We propose that this super-Eddington 
radiation pressure combined with inner disk radiation pressure disrupts the inner disk diverting accretion 
flow into the vertical direction leading to jet formation. Thus high radiation pressure is a necessary but 
not necessarily sufficient condition for jet formation. On the horizontal branch, the decrease of intensity
and $L_{\rm ADC}$ indicates that $\dot M$ decreases; however the blackbody temperature continues to increase
and there are various ways of explaining this, such as $\dot M$ decreasing in the disk while an enhanced
accretion flow is still affecting the neutron star. Evidence is presented that the flaring branch consists 
of unstable nuclear burning on the neutron star given the constant $L_{\rm ADC}$ and the increase of $L_{\rm BB}$, 
and the good agreement between $\dot m$ at the soft apex and the theoretical $\dot m_{\rm ST}$ below 
which burning is unstable. However, we should add the caution that while we now have evidence that the flaring 
branch can be explained in this way in the three Cyg-like Z-track sources, i.e. in GX\th 340+0 (Church et al. 
2006), in GX\th 5-1 (present paper) and in Cyg\th X-2 (Church et al. 2007), we do not claim that this is so 
in the other group of Z-track sources: the Sco-like sources. Preliminary evidence suggests that these
are similar on the NB and HB, but that the FB may be different. 

Radio emission is detected at the hard apex and horizontal branch showing that jets are present (Penninx 1989; 
Hjellming \& Han 1995; Berendsen et al. 2000) where we find that radiation pressure is strongest and is 
non-existent on the FB and soft apex. It was, of course, previously suggested that radiation pressure 
may be important in jet formation (Bisnovatyi-Kogan \& Blinnikov 1977; Begelman \& Rees 1984) and that the conical
opening in the inner accretion disk may perform some degree of collimation (Lynden-Bell 1978).

The present results do not support the standard view that $\dot M$ increases monotonically in the direction
HB - NB - FB, but have $\dot M$ constant in flaring and increasing from the soft apex to hard apex. Moreover,
in the standard view the source becomes super-Eddington at the soft apex whereas the present work indicates
the opposite, that this happens at the hard apex.

By combining timing and spectral analysis we have shown that the variation of the kHz QPO along the Z-track 
is highly correlated with $f/f_{\rm Edd}$, and the major change in the QPO frequencies takes place when this 
ratio equals unity, and the radiation pressure becomes large. This suggests that the higher frequency kHz QPO 
is an oscillation at the inner edge of the disk, i.e. the edge formed by the action of radiation
pressure removing a large section of the unperturbed disk, and that the edge moves to larger radial positions 
when the radiation pressure increases. A major feature of kHz QPO is that they are essentially a feature of 
the horizontal branch, fading away at the upper end of the normal branch and this is consistent with our model
for the Z-track as the high radiation pressure of the neutron star revealed by spectral fitting is also a 
feature of the horizontal branch. Fig. 11 shows that when $f/f_{\rm Edd}$ has fallen to $\sim$0.5, kHz QPO 
are no longer seen. Thus there is an indication that the kHz QPO only exist when part of the inner disk is 
blown away. The smallest Keplerian radius we measure has a value of 18 km 
on the upper normal branch and if QPO continued to exist on the normal branch, the Keplerian 
radius would decrease further. However, as the radiation pressure falls 
there would be no truncation of the disk at an edge acting as a site for the oscillation, and this is
consistent with the lack of kHz QPO on the lower normal branch and flaring branch.

Thus there is a simple explanation for the known occurrence of the higher frequency kHz QPO at a
preferred radial position in the disk: that this is the disk edge, and the results suggest that 
the variation of frequency is caused by change in position of the edge.

The physics of the accretion flow in the inner disk is very uncertain (e.g. van der Klis 2000), and
probably the most important question is whether the disk in general in LMXB meets the neutron star or is
truncated at an innermost stable orbit, an Alfv\'en radius or a sonic point, and this is relevant to
the QPO models which invoke truncation. However, it is difficult to say whether truncation of the disk due to 
relativistic, magnetospheric or other effects actually takes place. The present work presents a simple 
observational result on the higher frequency QPO and does not directly say anything about the lower 
frequency oscillation, and it is not appropriate to comment on the detailed physics of the QPO models here.
However, the present work indicates that the inner disk is not in general truncated, but can become truncated
on the upper NB and HB. Disk truncation is due to high radiation pressure and {\it we do not need}
to invoke truncation because of an innermost stable orbit, or an Alfv\'en radius external to the neutron star
or a sonic point.

\thanks{
This work was supported in part by the Polish KBN grant KBN-1528/P03/2003/25, by the Polish Ministry of 
Higher Education and Science grant no. 3946/B/H03/2008/34 and by PPARC grant PPA/G/S/2001/00052.}

\end{document}